\documentstyle[multicol,prd,aps,graphicx,epsfig]{revtex}
\begin{document}
\draft
%%%%%%%%%%%%%%%%%%%%%%%%%%%%%%%%%%%%%%%%%%%%%%%%%%%%%%%%%%%%%%%%%%%%%
%%%%%%%%%%%%%%%%%%%%%         Title       %%%%%%%%%%%%%%%%%%%%%%%%%%%
%%%%%%%%%%%%%%%%%%%%%%%%%%%%%%%%%%%%%%%%%%%%%%%%%%%%%%%%%%%%%%%%%%%%%

      \title{
%\begin{flushright} {\small IFT-P.  \,\, gr-qc/0211053} \end{flushright}
             Do static sources respond to massive scalar particles
             from the Hawking radiation as uniformly accelerated ones 
             do in the inertial vacuum?
            }

%%%%%%%%%%%%%%%%%%%%%%%%%%%%%%%%%%%%%%%%%%%%%%%%%%%%%%%%%%%%%%%%%%%%%       
%%%%%%%%%%%%%%%%%%%%     Authors & Addresses  %%%%%%%%%%%%%%%%%%%%%%%
%%%%%%%%%%%%%%%%%%%%%%%%%%%%%%%%%%%%%%%%%%%%%%%%%%%%%%%%%%%%%%%%%%%%%

\author{ 
         J.\ Casti\~ neiras\footnote{E-mail address: jcastin@ift.unesp.br}, 
         I.\ P.\ Costa e Silva\footnote{E-mail address: ivanpcs@ift.unesp.br}
and
         G.\ E.\ A.\ Matsas\footnote{E-mail address: matsas@ift.unesp.br} 
        }
\address{Instituto de F\'\i sica Te\'orica, Universidade Estadual Paulista,
         Rua Pamplona 145, 01405-900, S\~ao Paulo, SP, Brazil}

\maketitle

%%%%%%%%%%%%%%%%%%%%%%%%%%%%%%%%%%%%%%%%%%%%%%%%%%%%%%%%%%%%%%%%%%%%%%%
%%%%%%%%%%%%%%%%%%%%%           Abstract         %%%%%%%%%%%%%%%%%%%%%%
%%%%%%%%%%%%%%%%%%%%%%%%%%%%%%%%%%%%%%%%%%%%%%%%%%%%%%%%%%%%%%%%%%%%%%%
\begin{abstract}
We revisit the recently found equivalence for the response of a static
scalar source interacting with a {\em massless} Klein-Gordon field
when the source is (i) static in Schwarzschild spacetime, in the Unruh
vacuum associated with the Hawking radiation and (ii) uniformly
accelerated in Minkowski spacetime, in the inertial vacuum, provided
that the source's proper acceleration is the same in both cases. It is
shown that this equivalence is broken when the massless Klein-Gordon
field is replaced by a {\em massive} one.
\end{abstract}

%%%%%%%%%%%%%%%%%%%%%%%%%%%%%%%%%%%%%%%%%%%%%%%%%%%%%%%%%%%
%!!!!!!!!!!!!!!!!!!!!  Introducao  !!!!!!!!!!!!!!!!!!!!!!!%
%%%%%%%%%%%%%%%%%%%%%%%%%%%%%%%%%%%%%%%%%%%%%%%%%%%%%%%%%%%
\begin{multicols}{2}
\narrowtext

It was recently shown that the response $R^S (r_0,M)$ of a static
scalar source with radial coordinate $r_0$ outside a Schwarzschild
black hole of mass $M$ interacting with {\em massless} scalar
particles of Hawking radiation (associated with the Unruh vacuum) is
exactly the same as the response $R^M (a_0) \equiv q^2 a_0/4\pi^2$ of
such a source when it is uniformly accelerated in the inertial vacuum
of Minkowski spacetime, provided that the source's proper acceleration
$a_0$ is the same in both cases~\cite{HMS2}. (Here, $q$ is a coupling
constant.) Now, according to the Fulling-Davies-Unruh (FDU)
effect~\cite{U,FD}, the inertial vacuum in Minkowski spacetime
corresponds to a thermal state as seen by uniformly accelerated
observers confined to the Rindler wedge. Thus, the equivalence above
can be rephrased by saying that {\em the response of a static scalar
source with some fixed proper acceleration coupled to a {\em massless}
scalar field is the same when it interacts either (i) with the Hawking
radiation associated with the Unruh vacuum in Schwarzschild spacetime
or (ii) with the FDU thermal bath in Rindler spacetime}.  This came as
a surprise because structureless static scalar sources can only
interact with zero-frequency field modes. Such modes probe the global
geometry of spacetime and are accordingly {\em quite} different in
Schwarzschild and Rindler spacetimes. Moreover, since the response in
Schwarzschild spacetime $R^S (r_0,M)$ was expected to depend on $r_0$
and $M$ separately, it is striking that these parameters should
combine themselves precisely so that $R^S (r_0,M) = q^2
a_0(r_0,M)/4\pi^2$, as found in Ref.~\cite{HMS2}. The fact that such
an equivalence is not trivial can be also seen by the fact that it is
not verified when (i) the Unruh vacuum is replaced by the
Hartle-Hawking vacuum~\cite{HMS2}, (ii) the black hole is endowed with
some electric charge~\cite{CM} or even (iii) when the massless
Klein-Gordon field is replaced by the electromagnetic one~\cite{CHM1}.
A deeper understanding of why such an equivalence in the response is
verified for {\em massless} Klein-Gordon fields is still
lacking. While it {\em may} prove to be just a remarkable coincidence,
we feel that the problem deserves further analysis.  

In this paper we
show that providing some mass to the Klein-Gordon field is enough to
break the equivalence. The main technical difficulty
associated with the field quantization in Schwarzschild spacetime is
related to the fact that the positive and negative frequency modes
used to expand the quantum field cannot be expressed in terms of known
special functions. Although for massless spin-0 and spin-1 fields
outside Reissner-Nordstrom black holes, the quantization of the
low-energy sector admits an analytic
treatment~\cite{HMS2,CM,CHM1} (see also \cite{CL}), 
this is not the case for massive fields, for which a numerical analysis 
turns out to be required. 
Throughout this paper, we adopt natural units $c=G=\hbar=k_B=1$
and spacetime signature $(+ - - -)$.

The Schwarzschild line element describing a black hole of
mass $M$ can be written as~\cite{W}
\begin{equation}
     ds^2 = f(r) dt^2 - f(r)^{-1} dr^2 - r^2 \left( d\theta^2 + \sin^2
        \theta d\varphi^2 \right) \;, \label{1}
\end{equation}
where $f(r) \equiv 1-2M/r$. For the region outside the event horizon,
i.e. for $r > 2M $, we have a one-parameter group of isometries
generated by the timelike Killing field $\partial_t$.

Let us now consider a free Klein-Gordon field $\Phi (x^\mu)$ with mass
$m$ in this background described by the action
\begin{equation}
  S = 
        \frac{1}{2}\int d^4x \sqrt{-g} \; 
        [\nabla^\mu \Phi \nabla_\mu \Phi - m^2 \Phi ^2] \;,
   \label{2}
\end{equation}
where $g \equiv {\rm det} \{ g_{\mu \nu} \}$.
In order to quantize the field, we look for a complete set 
of positive-frequency solutions of the Klein-Gordon equation, 
$(\Box  + m^2)u^{\alpha}_{\omega l {\rm m}} = 0$ in the form
\begin{equation}
     u^{\alpha}_{\omega l {\rm m}}(x^\mu) = 
        \sqrt{\frac{\omega}{\pi}} \frac{\psi^{\alpha}_{\omega l}(r)}{r} 
Y_{l {\rm m}}(\theta,\varphi) e^{-i\omega t}  \;,
  \label{3}
\end{equation}
where $l \ge 0$, $ {\rm m} \in [-l,l]$  and $\omega$ are the 
angular momentum and frequency quantum numbers, respectively. 
Because the Klein-Gordon 
equation is of second order, there will be in general two
independent sets of normalizable solutions, here chosen to be incoming modes 
(i) {\em from the horizon} and (ii) {\em from infinity}  labeled 
by $\alpha = {\rm I}$ and $\alpha = {\rm II}$, respectively. The factor 
$\sqrt{\omega / \pi}$ has been inserted for later convenience 
and $Y_{l{\rm m}}(\theta,\varphi)$ are the spherical harmonics.
As a consequence, $\psi^{\alpha}_{\omega l}(r)$ must satisfy
\begin{equation}
  \left[ 
     - f(r) \frac{d}{dr} \left( f(r) \frac{d}{dr} \right) + V_{\rm eff}(r) 
  \right] 
  \psi^{\alpha}_{\omega l}(r) =  \omega^2 \psi^{\alpha}_{\omega l}(r) \;,
   \label{4} 
\end{equation}
where the effective scattering potential  $V_{\rm eff}(r)$ is given by 
\begin{equation}
     V_{\rm eff}(r) = 
         \left( 
        1- 2M/r  
        \right) 
        \left( 
        2M/r^3 + l(l+1)/r^2 + m^2
        \right) \;.
   \label{4a} 
\end{equation}
We note that {\em close to} and {\em far away from} the horizon we
have $V_{\rm eff}(r) \approx 0$ and $V_{\rm eff}(r) \approx m^2$,
respectively.  Thus, the frequency of the modes $u^{\alpha}_{\omega l
{\rm m}}$ with $\alpha = {\rm I}$ and $\alpha = {\rm II}$ will be
constrained so that $\omega \geq 0$ and $\omega \geq m$,
respectively. Now, it is convenient to recast Eq.~(\ref{4}) in a
Schr\"odinger-like form.  For this purpose, we define a new
dimensionless coordinate $y\equiv r/2M$ and perform the change of
variable $y \to x \equiv y + \ln\left| y-1 \right| $, so that
Eq.~(\ref{4}) becomes
\begin{equation}
     \left[ 
     - \frac{d^2}{dx^2} + 4M^2 V_{\rm eff}[r(x)] 
     \right] \psi^{\alpha}_{\omega l}(x) 
= 
        4M^2 \omega^2 \psi^{\alpha}_{\omega l}(x) \;.
   \label{8}
\end{equation}

We can expand the scalar field operator $\hat{\Phi}(x^\mu)$
in terms of annihilation $\hat{a}^{\alpha}_{\omega l {\rm m}}$ and creation
${\hat{a}^{\alpha \dagger}_{\omega l {\rm m}}}$ operators as usual:
\begin{equation}
     \hat{\Phi}(x^\mu) = 
        \sum_{\alpha = {\rm I}, {\rm II}}
        \sum_{l=0}^{\infty} \sum_{{\rm m} =-l}^{l}
        \int_0^{\infty} d\omega 
        \left[ 
        u^{\alpha}_{\omega l {\rm m}}(x^\mu)
        \hat{a}^{\alpha}_{\omega l {\rm m}} 
        + H.c. 
        \right] \;,
   \label{5}
\end{equation}
where $u^{\alpha}_{\omega l {\rm m}}(x^\mu)$ are orthonormalized according 
to the Klein-Gordon inner product~\cite{BDF}. As a consequence, 
$\hat{a}^{\alpha}_{\omega l {\rm m}} $ 
and 
$\hat{a}^{\alpha \dagger}_{\omega l {\rm m}} $ 
satisfy 
$ 
\left[
\hat{a}^{\alpha}_{\omega l {\rm m}},
{\hat{a}^{\alpha' \dagger}_{\omega' l' {\rm m}'}} 
\right] 
= \delta_{\alpha \alpha'} \delta_{l l'} 
\delta_{{\rm m} {\rm m}'} \delta(\omega - \omega')
$ 
and the Boulware vacuum $|0\rangle$ is
defined by 
$
\hat{a}^{\alpha}_{\omega l{\rm m} }|0\rangle = 0
$ 
for every $\alpha, \omega, l$ and ${\rm m}$~\cite{Bo}.

%%%%%%%%%%%%%%%%%%%%%%%%%%%%%%%%%%%%%%%%%%%%%%%%%%%%%%%%%%%
%!!!!!!!!!!!!!!!!!!!!  Quantizacao !!!!!!!!!!!!!!!!!!!!!!!%
%%%%%%%%%%%%%%%%%%%%%%%%%%%%%%%%%%%%%%%%%%%%%%%%%%%%%%%%%%%

Now, let us consider a pointlike static scalar source 
lying at 
$(r_0, \theta_0, \varphi_0)$ and described by  
\begin{equation}
     j(x^\mu) = (q/\sqrt{-h\;}) \delta(r-r_0)
     \delta(\theta-\theta_0) \delta(\varphi-\varphi_0) \;, 
\label{10}
\end{equation}
where $q$ is a small coupling constant and $h =- f^{-1} r^4 \sin^2
\theta$ is the determinant of the spatial metric induced on the
equal time hypersurface $\Sigma_t$. We will be interested in
analyzing the behavior of this source, coupled to the 
Klein-Gordon field $\hat{\Phi} (x^\mu)$ via the interaction action
\begin{equation}
  \hat{S}_I = \int d^4x \sqrt{-g}\; j\; \hat{\Phi} \;,
  \label{12}
\end{equation}
when it is immersed in the Hawking radiation emitted from the black
hole. All the calculations will be carried out at the tree level.

The total response, i.e., particle emission and
absorption probabilities per unit proper time of the source, is given by
\begin{equation}
     R^S \equiv  
       \sum_{\alpha = {\rm I}, {\rm II}}
       \sum_{l=0}^{\infty} 
       \sum_{{\rm m} =-l}^{l} 
       \int_0^{+\infty} d\omega  R^{\alpha}_{\omega l{\rm m}}  \;  ,
   \label{13}
\end{equation}
where
\begin{equation}
     R^{\alpha}_{\omega l{\rm m}} \! \equiv  \tau^{-1} \! \left\{ |{{\cal
     A}^{\alpha}_{\omega l{\rm m}}}^{\rm em}|^2 [1+ n^\alpha (\omega)] + |{{\cal
     A}^{\alpha}_{\omega l{\rm m}}}^{\rm abs} |^2 n^\alpha (\omega) \right\} 
\label{14}
\end{equation}
and $\tau$ is the total proper time of the source. Here $ {{\cal
A}^{\alpha}_{\omega l{\rm m}}}^{\rm em} \equiv \left\langle \alpha
\omega l {\rm m} \left| \hat S_I \right| 0 \right\rangle $ and $
{{\cal A}^{\alpha}_{\omega l{\rm m}}}^{\rm abs} \equiv \left\langle 0
\left| \hat S_I \right| \alpha \omega l {\rm m} \right\rangle $ are
the emission and absorption amplitudes, respectively, of Boulware
states $|\alpha \omega l {\rm m} \rangle$.  For the sake
of comparison with the equivalence found in the massless case, we
shall assume the Unruh vacuum, where $n^{\rm I} (\omega) \equiv (
e^{\omega\beta } - 1 )^{-1}$ and $n^{\rm II} = 0$, which
corresponds~\cite{BDF} to a thermal flux radiated away from the
horizon at temperature $\beta ^{-1} = 1/8\pi M $~\cite{HaU} as
measured by asymptotic static observers.

Structureless static sources~(\ref{10}) can only interact
with {\em zero-energy} modes and thus the response of the source 
in the Boulware vacuum vanishes. However, in the 
presence of a background thermal bath, the absorption and stimulated 
emission rates will lead in general to a non-zero response.  
In order to deal with zero-energy modes,
we need a ``regulator'' to avoid the appearance of intermediate
indefinite results (for a more comprehensive discussion on the
interaction of static sources with zero-energy modes, see
Ref.~\cite{HMS}). For this purpose, we let the coupling constant $q$
oscillate with frequency $\omega_0$ by replacing $q$ by $q_{\omega_0}
\equiv \sqrt{2} q \cos(\omega_0 t)$ in Eq.~(\ref{10}) and taking the
limit $\omega_0 \to 0$ at the end of our calculations. The
factor $\sqrt{2}$ has been introduced to ensure that the time average
$\left\langle |q_{\omega_0}(t)|^2\right\rangle_t = q^2$ since the 
absorption and emission rates are functions of $q^2$. 
Other equivalent regularization procedures can be devised~\cite{DS}. 
A straightforward calculation~\cite{CM} gives
\begin{equation}
R^S(r_0,M) = \frac{q^2 f(r_0)^{1/2}}{16 \pi^2 M r^2_0} \lim _{\omega_0
\to 0} \sum_{l=0}^{\infty} (2l + 1) |\psi^{\rm I}_{\omega_0 l}(r_0)|^2 
\;,
\label{response}
\end{equation}
where we have used the summation formula for spherical harmonics~\cite{GR}
$
\sum_{{\rm m} =-l}^{l} \left|Y_{l{\rm m}}(\theta_0,\varphi_0)\right|^2 =
(2l+1)/4\pi .
$
Note that only $\alpha = {\rm I}$ appears in Eq.~(\ref{response}).
This can be seen as reflecting the fact that the Unruh vacuum
corresponds to a thermal flux of particles being radiated only from
the horizon. It should be noticed, however, that the same response
(\ref{response}) holds when we replace the Unruh with the
Hartle-Hawking vacuum.  This is so because the extra
thermal flux coming from infinity in the Hartle-Hawking vacuum (which
should be considered in addition to the one coming from the horizon)
is composed of particles with frequency $\omega \geq m$ [see
discussion below Eq.~(\ref{4a})], i.e., this extra thermal flux is not
populated with zero-energy particles which are the only ones which can
interact with our source. 

In order to compute $R^S$ in Eq.~(\ref{response}), we shall evaluate
$\psi^{\rm I}_{\omega_0 l}$ (with $\omega_0 \to 0$) numerically. 
By using Eq.~(\ref{8}), it is easy to see that $\psi^{\rm I}_{\omega l}(x)$
has the following asymptotic forms when $ \omega < m$: 
\begin{equation}
       \! \psi^{\rm I}_{\omega l}(x) \approx \!
        \left\{ \!
          \begin{array}{lc}
           A_{\omega l} \left( 
                         e^{2iM\omega x} + {\cal R}_{\omega l} e^{-2iM\omega x} 
                        \right) 
           & \! (x>0, |x| \ll 1)  \\
           B_{\omega l}\, e^{-2M\sqrt{m^2 - \omega^2} x}   
           & \! (x \gg 1) ,
          \end{array} 
        \right.   
   \label{9}
\end{equation}
where $A_{\omega l}$ and $B_{\omega l}$ are appropriate normalization
constants. $|{\cal R}_{\omega l}|^2 =1$ is the reflection coefficient,
which is calculated by using Eq.~(\ref{9}) in Eq.~(\ref{8}). Indeed,
this solution describes modes that leave from the
horizon, ``scatter off the geometry'' and fall back to the horizon. 
The normalization constant $A_{\omega l} = (2\omega)^{-1}$ is analytically 
obtained (up to an arbitrary phase) by demanding that the normal 
modes~(\ref{3}) be orthonormalized with
respect to the Klein-Gordon inner product  (see, e.g.,
Ref.~\cite{HMS2} for details).  The modes $\psi^{\rm I}_{\omega l}$
can be obtained numerically for small $\omega/m$ and different $l$ 
values by evolving Eq.~(\ref{8}) with the effective 
potential~(\ref{4a}) and ``boundary conditions''~(\ref{9}).
The corresponding total response $R^S$ can be obtained, then, 
from Eq.~(\ref{response}). We note that the larger the
value of $l$, the higher the barrier of the scattering potential
$V_{\rm{eff}}(r)$~\cite{CCMV} and therefore the main contributions
come from modes with small $l$. How far we must sum over $l$
in Eq.~(\ref{response}) to obtain a satisfactory numerical result
will depend on how close to the black hole horizon the 
source lies. The closer to the horizon the further over $l$  
we must sum. We have checked our numerical code 
for $m=0$ where the response is known analytically~\cite{HMS2}.

Our results for $R^S$ will be exhibited in comparison with the
response $R^M$ obtained when our scalar source is uniformly
accelerated (with proper acceleration $a_0$) in the usual inertial
vacuum in Minkowski spacetime. $R^M$ can be equivalently computed
either with respect to inertial or uniformly accelerated observers.
We favor the latter here. Accordingly, we shall perform the
quantization of the massive Klein-Gordon field in the Rindler wedge,
which can be described by the line element \cite{BDF}
\begin{equation}
     ds^2 = 
        e^{2a_0 \xi}\left(d\tau ^2 - d\xi^2 \right ) - dx^2 - dy^2 
  \label{17} 
\end{equation}
with $-\infty <\tau, \xi, x, y < +\infty $.
The corresponding Klein-Gordon orthonormalized positive-frequency
modes are 
\begin{eqnarray}
u_{\omega {\bf{k}}_\bot} (x^{\mu}) 
&=& 
\frac{\sqrt{ \sinh (\pi \omega / a_0)}}{2 \pi^2 a_0^{1/2}} 
K_{i \omega /a_0 }\left[ \frac{({\bf{k}}_\bot^2 + m^2)^{1/2} e^{a_0 \xi}
}{a_0 } \right] \nonumber \\ 
& \times & 
e^{i {\bf{k}}_\bot \cdot {\bf{x}}_\bot - i\omega \tau}  \;,
\label{18} 
\end{eqnarray}  
where 
$K_{\nu}(x)$ 
is the modified Bessel function \cite{GR}, 
${\bf{x}}_\bot = (x,y)$
and 
${\bf{k}}_\bot \equiv (k_x,k_y)$ 
denotes the momentum transverse to the direction of acceleration. 
In these coordinates, our source with constant proper acceleration
$a_0$ will be described by 
$ 
j(x^\mu) = q \delta(\xi) \delta(x) \delta(y) 
$.
The total response is given, in this case, by
\begin{equation}
     R^M \equiv  
       \int d {\bf k}_\bot^2       
       \int_0^{+\infty} d\omega \, R_{\omega {\bf k}_\bot}  \;  ,
   \label{13.3}
\end{equation}
where
$$
R_{\omega {\bf k}_\bot} \! 
\equiv  
\tau^{-1} \! 
\left\{ |{{\cal A}_{\omega {\bf k}_\bot}^{ \rm em}}|^2  
[1+ n (\omega)] + 
|{{\cal A}_{\omega {\bf k}_\bot}^{\rm abs}}|^2 
n (\omega) 
\right\} .
$$
Here 
$ {{\cal A}_{\omega {\bf k}_\bot}^{\rm em}}
\equiv 
\left\langle \omega {\bf k}_\bot\left| \hat S_I \right| 0 \right\rangle
$ and 
$ {{\cal A}_{\omega {\bf k}_\bot}^{ \rm abs}}
\equiv 
\left\langle 0 \left| \hat S_I \right| \omega {\bf k}_\bot \right\rangle
$ are the emission and absorption amplitudes, respectively, of
Rindler states $| \omega {\bf k}_\bot \rangle$ and 
$n(\omega) = 1/ (\exp(\beta \omega) -1)$, where $\beta ^{-1} = a_0/2\pi$
is the temperature of the FDU thermal bath associated with the 
inertial vacuum as measured by the Rindler observer lying 
at $\xi = 0$. The response can be shown to be
\begin{equation}
       R^M(a_0) = \frac{q^2 a_0}{2 \pi^2} \int_{0}^{\infty}dx~x
         K^2_0 \left( \sqrt{x^2 + (m/a_0)^2} \right) \;.
\label{20}
\end{equation}

\begin{figure}
 \begin{center}
 \mbox{\epsfig{file=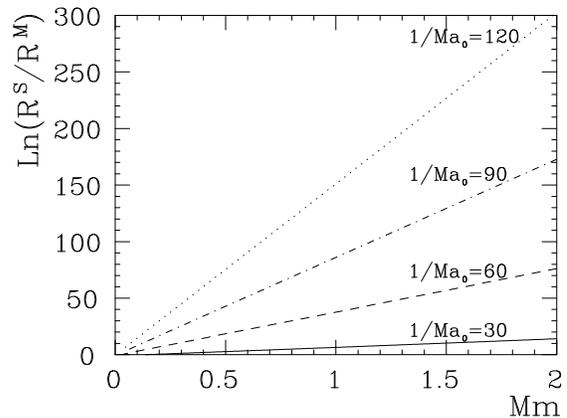,width=0.36\textwidth,angle=90}}
 \end{center}
\caption{ ${\rm Ln}(R^S/R^M)$ is plotted as a function of $m$ (where,
for the sake of convenience, we have used the black hole mass $M$ as a
standard scale).  The equality between $R^S$ and $R^M$ is recovered
when $m \to 0$, as expected, but not for general
values of $m$. [In plotting this graph, we have summed up to $l=18$ in
Eq.~(\ref{response}).]
}
\label{R^Sxm}
\end{figure}

In Fig.~\ref{R^Sxm} we plot $R^S (a_0, M)/R^M (a_0)$ as a function of
the scalar field mass $m$ for different $a_0$'s. We recall that the
source's proper acceleration $a_0$ is a one-to-one function of its
radial position $r_0$: $ a_0 = M/(r_0^2 \sqrt{1-2M/r_0 \,}).  $ We
note that in general $R^S (a_0, M) \neq R^M (a_0)$. The equivalence is
only recovered when $m \to 0$.

\begin{figure}
\begin{center}
\mbox{\epsfig{file=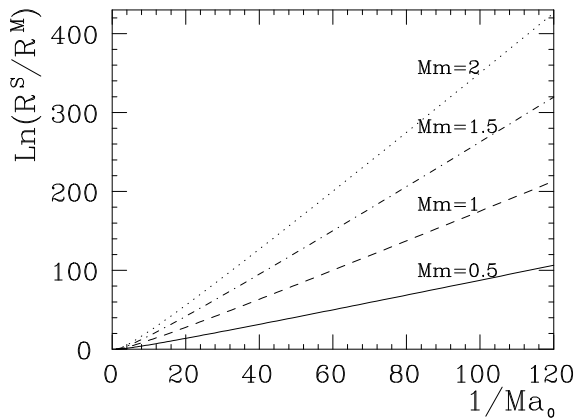,width=0.36\textwidth,angle=90}}
\end{center}
\caption{In this Figure, it becomes clear how the equality between
$R^S$ and $R^M$ is broken for various masses of the Klein-Gordon
field. Note that the more we move the source away from the horizon
(which corresponds to decreasing $a_0$), the more $R^S$ deviates from
$R^M$. On the other hand, $R^S \to R^M$ as we approach the horizon 
($a_0 \to \infty$) (see discussion in the text). [In plotting this 
graph, we have summed up to $l=18$  in Eq.~(\ref{response}).]}
\label{R^Sxa}
\end{figure}

In Fig.~\ref{R^Sxa} we display how the equivalence is broken for various
values of $m$. $R^S$ deviates from $R^M$ as one goes
away from the horizon at a rate depending on $m$.
Note that indeed $R^S$ approaches $R^M$ as $m$ goes to zero.
The fact that $R^S/R^M \to 1$ in the massive case
for $a_0 \to \infty$ could be analytically predicted from the 
equivalence found in the massless case as follows. On the one hand, 
close to the black hole horizon, the potential (\ref{4a}) is not 
significantly influenced by the mass of the field. 
As a consequence, in this region, massive and non-massive 
outgoing modes $\psi^{\rm I}_{\omega l}$, which
are the relevant ones in Eq.~(\ref{response}), behave similarly.
Thus, it is expected that for sources close enough to the 
horizon, i.e. $a_0/m \gg 1$, $R^S$ should be approximately 
the same for massive and non-massive fields.
On the other hand, it is easy to see from Eq.~(\ref{20}) that 
the same statement is true for $R^M$. Hence the equivalence found
for the massless case also implies that $R^S/R^M \to 1$ for 
{\em massive} fields as long as $a_0/m \to \infty$.

We have shown in this paper that the response of a static 
scalar source with some fixed proper acceleration minimally coupled to 
a {\em massive} scalar field is not the same when it interacts 
(i) with the Hawking radiation in Schwarzschild spacetime 
and (ii) with the FDU thermal bath in Rindler spacetime.
This emphasizes how unexpected was the equivalence 
found in the massless case for the behavior of a classical source 
in quite different spacetimes with respect to an observable 
(the response) which depends on the {\em global} structure of 
the underlying background.

\begin{flushleft}
{\bf{\large Acknowledgments}}
\end{flushleft}

J.C. and I.S. would like to acknowledge full support from Funda\c
c\~ao de Amparo \`a Pesquisa do Estado de S\~ao Paulo. G.M. is
thankful to Conselho Nacional de Desenvolvimento Cient\'\i fico e
Tecnol\'ogico for partial support.

%%%%%%%%%%%%%%%%%%%%%%%%%%%%%%%%%%%%%%%%%%%%%%%%%%%%%%%%%%%%%%%%%%%%%%%%
%                               BIBLIOGRAPHY
%%%%%%%%%%%%%%%%%%%%%%%%%%%%%%%%%%%%%%%%%%%%%%%%%%%%%%%%%%%%%%%%%%%%%%%%

\end{multicols}
\end{document}